\documentclass[showpacs,showkeys,aps,prb,twocolumn,superscriptaddress,citeautoscript]{revtex4}

\usepackage{graphicx}
\usepackage{pstricks}

\newcommand{\ignore}[1]{}
\newcommand{\etal}{{\it et al.}}

\newcommand{\vG}{\mathbf{G}} % symbol for vector G
\newcommand{\vk}{\mathbf{k}} % symbol for vector k
\newcommand{\Gk}{\vk+\vG}
\newcommand{\Gkp}{\vk+\vG'}

\begin{document}

\title{Screening in 2D: \textit{GW} calculations for surfaces and thin films using
the repeated-slab approach}
\author{Christoph Freysoldt}
%\email{freyso@fhi-berlin.mpg.de}
\author{Philipp Eggert}
\author{Patrick Rinke}
\affiliation{Fritz-Haber-Institut der Max-Planck-Gesellschaft, Faradayweg 4--6, 14195 Berlin, Germany}
\author{Arno Schindlmayr}
\affiliation{Fritz-Haber-Institut der Max-Planck-Gesellschaft, Faradayweg 4--6, 14195 Berlin, Germany}
\affiliation{Institut f\"ur Festk\"orperforschung, Forschungszentrum J\"ulich, 52425 J\"ulich, Germany}
\affiliation{Department Physik,
Universit\"at Paderborn, 33095 Paderborn, Germany}
\author{Matthias Scheffler}
\affiliation{Fritz-Haber-Institut der Max-Planck-Gesellschaft, Faradayweg 4--6, 14195 Berlin, Germany}

\begin{abstract}
In the context of photoelectron spectroscopy, the $GW$ approach has developed into the 
method of choice for computing excitation spectra of weakly
correlated bulk systems and their surfaces.
To employ the established computational schemes that have been developed
for three-dimensional crystals, two-dimensional systems
are typically treated in the repeated-slab approach.
In this work we critically examine this approach and identify three important aspects for
which the treatment of long-range screening in two dimensions differs from the bulk:
(1) anisotropy of the macroscopic screening
(2) {\bf k}-point sampling parallel to the surface
(3) periodic repetition and slab-slab interaction.
For prototypical semiconductor (silicon)
and ionic (NaCl) thin films we quantify the individual contributions of 
points~(1) to (3)
and develop robust and efficient correction schemes derived from 
the classic theory of dielectric screening.
\end{abstract}

%\keywords{repeated-slab approach, $G_0W_0$ quasiparticle calculations,
%dielectric screening, surfaces, thin films}

\maketitle
\section{Introduction}

Surface science has developed into a highly
active and multidisciplinary area of research that has produced a
rich variety of surface-sensitive experimental techniques. 
In this context theory and calculations have become an invaluable tool
for interpreting the often complex and indirect data and for making
predictions for
structures or properties still inaccessible experimentally.
 
For many experimentally and technologically relevant
questions the electronic structure of a system is of central importance.
Experimentally, it can be probed, for example, by photoemission 
spectroscopy.
%The steady improvement in the experimental energy and momentum resolution 
%requires corresponding refinements in the theoretical tools
%to interprete the experimental spectra.
Theoretically, many-body perturbation theory in the
$GW$ approximation, where $G$ is the Green's function
and $W$ is the dynamically screened Coulomb interaction,
has developed into the method of choice for describing electronic
excitations as measured in direct and inverse photoemission in 
weakly correlated solids and their surfaces
\cite{Hedin,Aulbur,Rinke,GWsurfaces,InsulatorImageStates,benzeneGraphite,
GWforDefectsPRL}.

\begin{figure}
\includegraphics[width=0.95\columnwidth,clip]{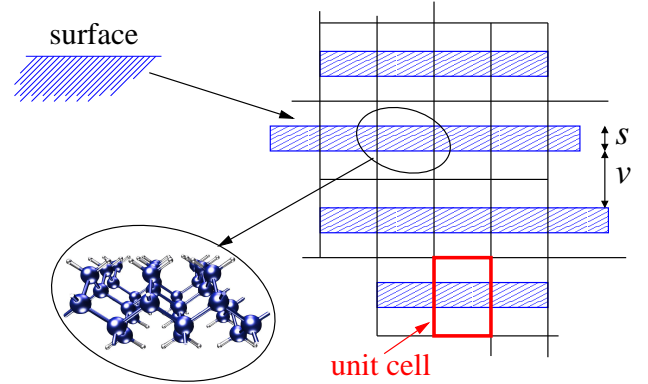}
\caption{Repeated-slab approach for surfaces and thin films schematically, with $s$ being
the slab thickness and $v$ the slab separation.}
\label{fig:repslab}
\end{figure}

Most $GW$ implementations, notably such that employ pure or augmented
plane-waves as basis functions \cite{HybertsenLouie,gwstPRL,Lebegue,Kotani},
but also others \cite{Rohlfing},
rely on three-dimensional periodic boundary conditions,
which are appropriate for crystalline bulk systems.
To treat systems with a reduced periodicity like surfaces, thin films,
nanowires, clusters, or molecules, three-dimensional periodicity 
is imposed artificially.
For this, the system of interest is placed in a three-dimensional unit cell,
with an empty (vacuum) region to separate the physical system from its 
periodic images in the broken-symmetry direction(s); 
see Fig.~\ref{fig:repslab}. This procedure is frequently referred to as 
``supercell approach,'' but for systems with a two-dimensional periodicity 
we prefer the more descriptive term ``repeated-slab approach.''

% - briefly review surface applications
%   a) using model self energies
%   b) using ab initio, but not tested systematically
In this paper we will focus on the two-dimensional case, i.e., 
the description of surfaces and thin films. 
Already in the early days of modern $GW$ calculations,
Hybertsen and Louie applied the then new methodology to simple semiconductor
surfaces using the repeated-slab approach \cite{HybertsenLouieSurfaces}.
In their paper they raised several fundamental and technical questions.
Of importance here is their speculation that
``the crucial change in the self-energy operator at the surface may
be largely contained in the Green's function,'' while ``the screened
interaction may more closely follow the variation in the local density.''
They further emphasize that their preliminary conclusions ``will require
further examples {\ldots} as well as a critical
evaluation of the slab approach for the surface self-energy operator''.
Surprisingly, these issues have only rarely been addressed
in later studies and the ``critical evaluation of the slab approach''
is still missing in the literature.
This paper is a contribution to fill this gap, illustrating that the
screened interaction does \emph{not} simply follow the local density
but unfortunately is substantially influenced by the repeated-slab geometry.

In principle it is straightforward to investigate how the artificial
periodic repetition (in one or more dimensions) affects a given
computational method by increasing
the vacuum region until the properties of interest exhibit no dependence
on this parameter anymore.
This is easily achieved in density functional theory (DFT) with the commonly
employed local or semilocal functionals, provided that appropriate correction
schemes for the low-order electrostatic multipole moments are applied
\cite{DFTdipoleCorrection}, if necessary. 
The decoupling becomes more difficult
for $GW$ due to the long-range nature of the (screened) Coulomb interaction,
as has been explicitly shown for a Na$_4$ cluster \cite{Na4}.
%A static correction scheme similar to the ground-state
%case does not exist, as the corresponding multipoles are
%induced only in the presence of the additional electron or hole and hence
%not known explicitly.
To solve the decoupling problem, it has been proposed to cut off the Coulomb
interaction
in the broken-symmetry directions \cite{Na4,LouieNanoWire,Rozzi,Sohrab}.
However, to ensure that no interactions within the physical system are
truncated, the vacuum region must be at least as large as the physical
system in these schemes, which makes the approach unproportionally expensive
(in terms of the computational effort) for thicker slabs.
Moreover, the $\mathbf k$-point sampling
of the Brillouin zone, the computational parameter naturally associated
with these long-range effects, becomes a critical convergence parameter
along the nontruncated directions \cite{LouieNanoWire,Sohrab}. 
As we will show below, even this
seemingly technical issue is related to the screening and
hence the physical properties of the system.

An obvious way to avoid these spurious interactions for systems with broken
translational symmetry is to abandon the concept of 
periodic boundary conditions altogether in the relevant directions
and perform the calculations entirely in real space.
For semi-infinite jellium surfaces such a $GW$ embedding scheme has been
successfully implemented \cite{Fratesi1,Fratesi2}. 
Its extension to realistic
surfaces, however, is computationally still too expensive. 
A real-space implementation for finite systems has also been
reported \cite{Tiago,Tiago2}, but its applicability to systems with periodicity
in one or more directions remains to be shown.

% purpose of this work:
% a) investigate the nature and the magnitude of the effects introduced
% by the periodic repetition
% b) identify computational parameters that cannot be transfered from
%    bulk $GW$ or slab DFT calculations and require independent tests
% c) find out whether the ''traditional'' repeated-slab approach can
%    be used in practice
The purpose of this work is to reevaluate the performance of the
repeated-slab approach when no
Coulomb-truncation technique is employed. We address
the nature and the magnitude of the effects introduced by the
periodic repetition and propose a robust correction scheme
that allows to extract the isolated-slab limit already
from very small vacuum separations.
To achieve this, it proved necessary to explicitly take the anisotropy
of the macroscopic screening into account.
In addition, we have found that the $\mathbf k$-point sampling
of the Brillouin zone becomes a more critical parameter 
for $GW$ repeated-slab calculations than recognized so far.
Appropriate samplings sufficient
for bulk $GW$ or slab DFT calculations cannot be
transferred to surfaces in general.

For weakly correlated bulk systems $GW$ calculations are
now routinely performed. Remaining open issues, such as
computational efficiency, influence of pseudopotentials
or self-consistency, are actively being addressed
\cite{gwstPRL,Rohlfing,Tiago,Rinke,FriedrichLAPW,Bruneval,KresseGW,SchilfgaardeSCGW,Delaney}
but do
not affect the conclusions drawn in this paper. % TODO: Zitate
For more strongly correlated systems, it becomes necessary
to go beyond $GW$,
but these schemes often include the $GW$ self-energy diagrams as lowest
order \cite{Biermann,Sun}.
Similarly, the Bethe-Salpeter approach to electron-hole excitations, as 
probed in optical absorption or electron energy-loss spectroscopy, builds 
on the $GW$ self-energy \cite{Onida}.
In these schemes, screening plays a similar role as for $GW$ 
calculations, but they go beyond the scope of the present work. 

We note also that the surface electronic structure of a
slab may differ from that of a semi-infinite solid. Surface resonances and
plasmons, for example, may be affected by confinement effects. The finite
thickness of the slab becomes an additional potential source of error in
surface calculations, but will not affect free-standing films or
heterostructures. Converging the surface electronic structure
with respect to slab thickness and/or developing robust correction
schemes is a separate problem that would require a detailed study in
its own right. We will briefly address the relevant issues in 
Sec.~\ref{sec:discussion}.

The remainder of this paper is organized as follows: In 
Sec.~\ref{sec:compute} we present the methodological and physical
aspects of the repeated-slab approach. After briefly summarizing the $GW$ 
space-time method (Sec.~\ref{sec:gwst}),
we focus first on the anisotropy of
macroscopic screening (Sec.~\ref{sec:aniso})
and the $\mathbf k$-point convergence parallel to the surface 
(Sec.~\ref{sec:park}). Then we discuss the influence of the
periodic repetition and the associated convergence of
the band energies with respect to the 
vacuum separation between the slabs (Sec.~\ref{sec:vac}).
In Sec.~\ref{sec:discussion} we summarize our main findings
and put our conclusions
into the context of other computational $GW$ schemes.
In the Appendix %\ref{app:imgch}
we present the computational 
scheme for the classical dielectric models employed.

\section{Long-range screening}
\label{sec:compute}

\subsection{The \textit{GW} space-time method}
\label{sec:gwst}
All calculations have been performed with the $GW$ space-time method
\cite{gwstPRL,Rieger,gwstGL}. Due to its advantageous linear scaling behavior
with respect to the $\mathbf k$-point sampling of the Brillouin zone
it is ideally suited for performing the extensive convergence studies
presented below.
We have recently extended the code to include
the anisotropy in the long-range screening \cite{myCPC},
a crucial point for the repeated-slab approach, as we will demonstrate
below. In the following, we will only briefly sketch out the computational
scheme and refer the interested reader to the previous papers for
further details.

In the space-time method, the Green's function $G$ is
constructed in real space ($\bf r$ and $\bf r'$) 
and imaginary time ($i\tau$)
from the Kohn-Sham wave functions $\phi_{n\mathbf k}$ and energies 
$\epsilon_{n\mathbf k}$ (the Fermi level is set as the energy zero).
\begin{equation}
G(\mathbf r, \mathbf r ';\pm i\tau) = 
 \frac{\mp i \Omega}{(2\pi)^3}\int\limits_{\rm BZ}\!d^3 k
\sum\limits_n \varphi_{n \mathbf k}(\mathbf r) \varphi_{n \mathbf k}^*( \mathbf r')
e^{- \epsilon_{n{\mathbf k}}\tau}
,
\end{equation}
where $\Omega$ denotes the unit-cell volume and the integral over
$\bf k$ runs over the first Brillouin zone. Depending on the signs taken, 
the state summation over $n$ runs over unoccupied (occupied) states
for positive (negative) imaginary times.

The polarizability is calculated in the random-phase approximation by
\begin{equation}
P(\mathbf r ,\mathbf r '; i\tau) = -2 i 
G(\mathbf r ,\mathbf r '; i\tau) 
G(\mathbf r,\mathbf r ';- i\tau)
\label{eq:defP}
\end{equation}
and is then Fourier transformed to reciprocal space and imaginary 
frequency ($i\omega$).
The polarizability $P$ is used to compute the screened interaction
$W$ via the symmetrized dielectric matrix 
\begin{equation}
\tilde \varepsilon_{\vk}(\vG,\vG';  i\omega) = \delta_{\vG\vG'} - 
\frac{4\pi}{|\Gk||\Gkp|} P_{\vk}(\vG,\vG'; i\omega)
\; .
\label{eq:eps_form}
\end{equation}
From its matrix inverse for each $\mathbf k$ and 
$i\omega$, the screened interaction
\begin{equation}
W_{\vk}(\vG,\vG';  i\omega) = 
\frac{4\pi}{|\Gk||\Gkp|} \tilde\varepsilon^{-1}_{\vk}(\vG,\vG';  i\omega)
\label{eq:define_W}
\end{equation}
is obtained and then Fourier transformed back to real space and imaginary
time. The self-energy is then computed as
\begin{equation}
\Sigma(\mathbf r , \mathbf r ';  i\tau) =  i G(\mathbf r , \mathbf r ';  i\tau) W(\mathbf r , \mathbf r ';  i\tau)
\; .
\label{eq:sig}
\end{equation}
To obtain the self-energy corrections to the Kohn-Sham energies
$\epsilon_{n\mathbf k}$, the
matrix elements of the perturbation operator $\Sigma - V_{\rm xc}$ 
(where $V_{\rm xc}$ denotes the local exchange-correlation potential)
are computed and Fourier transformed to the imaginary frequency axis.
These matrix elements are then analytically continued to the real 
frequency axis by fitting a multi-pole function
on the imaginary frequency axis \cite{Rieger}.
Finally, the quasiparticle energies $\epsilon^{\rm qp}_{n\vk}$ are
given by the solution of
\begin{equation}
\label{eq:qpe}
\epsilon^{\rm qp}_{n\vk} = \epsilon_{n\vk} + 
\langle\varphi_{n\vk}|
\Sigma(\epsilon^{\rm qp}_{n\vk})-V_{\rm xc}
|\varphi_{n\vk}\rangle
\; ,
\end{equation}
where $\langle\varphi_{n\vk}||\varphi_{n\vk}\rangle$ denotes matrix elements with respect
to the Kohn-Sham wave functions.

\subsection{Anisotropy of the macroscopic screening}
\label{sec:aniso}

An important point for computing the screened interaction in reciprocal
space [Eq.~(\ref{eq:eps_form})--(\ref{eq:define_W})] is the $1/k^2$ singularity
of the Coulomb potential as $\mathbf k\rightarrow \mathbf 0$. For the 
$\mathbf G=\mathbf G'=\mathbf 0$ element of the symmetrized dielectric
matrix [Eq.~(\ref{eq:eps_form})] it is cancelled by 
the $k^2$ behavior of the polarizability \cite{metals}.
In practice, we enforce
this cancellation analytically by performing a Taylor expansion for
the polarizability around the $\Gamma$ point $\mathbf k=\mathbf 0$.
However, the expansion depends in general on the spatial direction
in which the $\Gamma$ point is approached.
This directional dependence reflects the anisotropy in the macroscopic
screening and introduces a nonanalytic, yet finite contribution
to the inverse dielectric matrix \cite{PickCohenMartin}.

\begin{figure}
\center
\includegraphics[scale=0.4,clip]{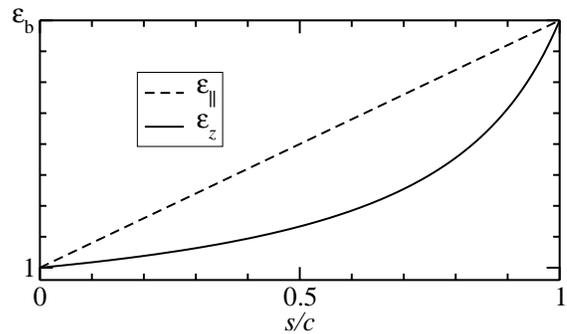}
\caption{Dielectric tensor for repeated-slab systems according
to effective-medium theory as a function of the $s/c$ ratio 
($s$ slab thickness, $c$ total size of the simulation box):
component $\varepsilon_\parallel$ parallel to the surface
and $\varepsilon_z$ perpendicular to it.}
\label{fig:effm-aniso}
\end{figure}

The screened interaction in the vicinity of the $\Gamma$ point then
takes the form
\cite{PickCohenMartin,Hott,myCPC}
\begin{equation}
W_{\mathbf k}(\mathbf 0, \mathbf 0; i\omega) 
\rightarrow \frac{4\pi}
{\mathbf k^{\rm T} \mathbf L(i\omega) \mathbf k}
\label{eq:wklk}
\; ,
\end{equation}
where $\mathbf L(i\omega)$ denotes the macroscopic dielectric tensor.
The tensor expression in the denominator 
reduces to a scalar when the macroscopic screening is isotropic.  
To our knowledge, most $GW$ implementations
explicitly exploit this simplification
(exceptions being the exact treatment of Ref.~ \onlinecite{Hott}
and the approximate but accurate\cite{myCPC} tensor treatment
of Ref.~ \onlinecite{Pulci})
and therefore implicitly assume
that the screening is isotropic.
However, this is not the case for repeated-slab systems,
as is easily demonstrated. Assuming that the slabs in 
Fig.~\ref{fig:repslab} are homogeneous dielectrics of finite
thickness $s$ with an isotropic bulk dielectric constant 
$\varepsilon_{\rm b}$ and separated by a vacuum region with thickness $v$,
the dielectric tensor components of the repeated-slab system
are given by
\begin{eqnarray}
\varepsilon_{\parallel} &=& \frac{\varepsilon_{\rm b} s + v}{s+v}
= 1 + (\varepsilon_{\rm b} - 1) \frac sc \label{eq:eps_par} \; ,\\
\varepsilon_z^{-1} &=& \frac{\varepsilon_{\rm b}^{-1} s + v}{s+v}
= 1 - (\varepsilon_{\rm b} -1) \frac{s}{\varepsilon_{\rm b} c} \label{eq:eps_z}
\; ,
\end{eqnarray}
where $c=s+v$ denotes the total height of the simulation cell.
$\varepsilon_{\parallel}$ (parallel to the surface)
and $\varepsilon_z$ (perpendicular to it) agree for isotropic 
systems ($s/c=0$ or $1$) but deviate considerably for $s/c$ ratios
between these limiting cases (cf.  Fig.~\ref{fig:effm-aniso}).
The ratio
$\varepsilon_\parallel / \varepsilon_z$ becomes largest for
$s/c=1/2$. In practice, the parameters of repeated-slab systems
are often close to this ratio of maximum anisotropy.

In the $GW$ calculations
the singular elements of $W$ require a special treatment 
in the subsequent computational steps, which formally involve an
integration over
the Brillouin zone. Since the singularity is integrable, the
result is formally well defined, but cannot be approximated by
a finite summation. The solution is to split the screened interaction
according to
\begin{equation}
W = W^{\rm lr} + W^{\rm sr}
\end{equation}
into a long-range part $W^{\rm lr}$ and a short-range remainder 
$W^{\rm sr}$.
For $W^{\rm lr}$, a simple analytic form is chosen that exhibits
the same singularity as $W$,
but for which the integral can be computed analytically.
\cite{Lebegue,HybertsenLouie,Rieger,Hott,Pulci,Wenzien,myCPC} 
The remainder $W^{\rm sr}$ then becomes nonsingular, and its integral
can safely be replaced by the sum over a finite $\mathbf k$-point grid.

\begin{figure}
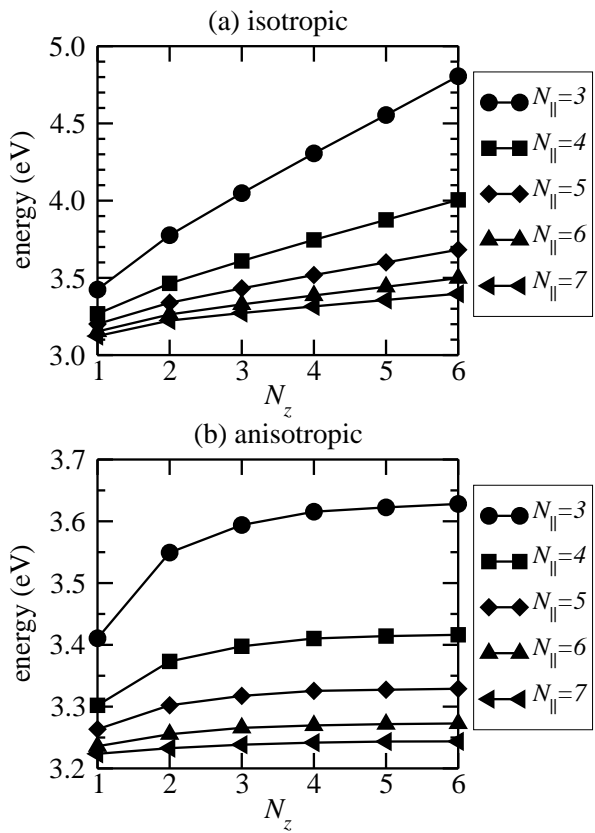

\begin{center}
\includegraphics[width=0.9\columnwidth,clip]{kconv-nz-avg.eps}

\includegraphics[width=0.9\columnwidth,clip]{kconv-nz.eps}
\end{center}
\caption{Convergence of the fundamental gap of a 
H-saturated four-layer Si(100) slab with respect to the number
of $\mathbf k$ points $N_z$
perpendicular to the surface for (a) the isotropic and (b) the 
anisotropic approach.
The quantitative behavior depends on the sampling in the 
parallel direction ($N_\parallel\times N_\parallel$).
Note the different scales of the two graphs.
}
\label{fig:kz-conv}
\end{figure}

We demonstrate the importance of an anisotropic treatment
for $W^{\rm lr}$ 
 for the case of a hydrogen-saturated
four-layer Si(100) slab with a vacuum region equivalent to four layers
of silicon. 
The calculated nonzero elements of this repeated-slab system's dielectric 
tensor are $\varepsilon_{xx}$=5.1, $\varepsilon_{yy}$=5.5,
and $\varepsilon_{zz}$=2.2 at the smallest imaginary frequency
$\omega$=0.036 hartree, highlighting the large anisotropy
predicted by the electrostatic considerations above.
In Fig.~\ref{fig:kz-conv}, we show the convergence of the 
fundamental gap\cite{remark:gap} with respect
to the $\mathbf k$-point sampling $N_z$ perpendicular to the surface.
It is obvious that isotropic averaging for the screened interaction,
i.e., using a scalar rather than a tensorial expression for the
singularity of $W^{\rm lr}$, leads to an 
unphysical linear increase in the band gap, which will not level
off when $N_z$ is increased further.
The reason for this linear divergence lies
in the inadequate treatment of the singularity in Eq.~(\ref{eq:wklk}), 
which is not fully removed \cite{myCPC}.
In contrast,
the anisotropic treatment converges rapidly.

Only the proper anisotropic treatment allows us to investigate
the importance of the $\mathbf k$-point sampling in the direction perpendicular
to the surface.
This convergence has not been discussed for repeated-slab systems before. 
Using the anisotropic treatment in the $GW$ space-time method and
a $N_\parallel\times N_\parallel \times N_z$ sampling, we
find that the self-energy corrections exhibit a $1/N_z$
behavior, the magnitude of which rapidly decreases with increasing
$N_\parallel$ \cite{myCPC}.
Such a behavior might result from the remaining approximations made for
the $\Gamma$ point. 
In practice, the convergence with respect to $N_z$ plays a relevant role
only for coarse $\mathbf k_\parallel$ samplings.
We will address convergence with respect to this parallel sampling in
Sec.~\ref{sec:park}. Test calculations indicate that the qualitative behavior
is not affected by the choice of $N_z$.  
For simplicity, we have therefore used $N_z=1$ for the calculations reported in 
the following.

\subsection{$\mathbf k$-point sampling parallel to the surface}
\label{sec:park}

\begin{figure}
\includegraphics[width=0.95\columnwidth,clip]{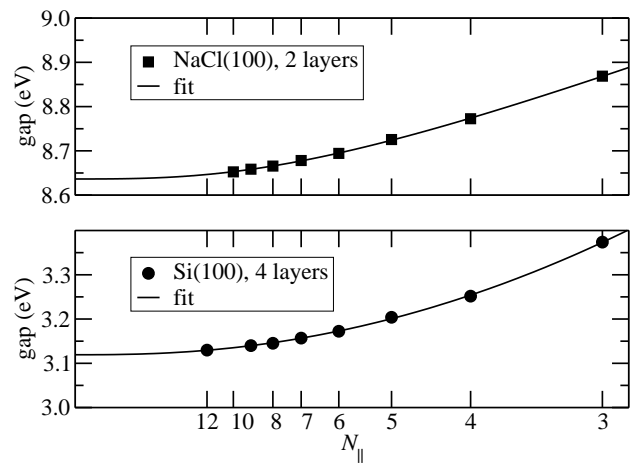}
\caption{Band gap for
a two-layer NaCl(100) slab and a hydrogen-saturated four-layer Si(100) slab
with a $N_{\parallel}\times N_{\parallel}\times 1$
$\mathbf k$-point sampling as a function of 1/$N_{\parallel}$.
The vacuum region is 20 bohr wide for both systems.
The fit function is explained in the text.}
\label{fig:kcurves}
\end{figure}
The $\mathbf k_\parallel$ sampling is 
a further critical aspect in $GW$ calculations for surfaces and thin films,
which may not have attracted sufficient attention so far.
We demonstrate the importance of this issue for two representative slab
systems:
a two-layer NaCl(100) slab and a hydrogen-saturated four-layer Si(100) 
slab.
In Fig.~\ref{fig:kcurves} the gap for a
$N_{\parallel}\times N_{\parallel}\times 1$ sampling is plotted
as a function of $1/N_{\parallel}$. 
In both cases we observe a similar shape of the curves
that does not follow a simple power law as observed for $N_z$. 
We will show in the following that this can be understood in terms of 
the particular decay behavior of the screened interaction
in a slab system and can be modeled using a simple analytic function.

\begin{figure}
\center
\includegraphics[width=0.95\columnwidth]{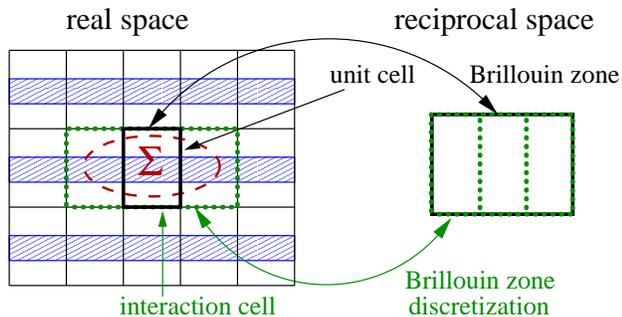}
\caption{The nonlocality of two-point functions, such as $\Sigma$, is defined
within the interaction cell. In the space-time method, its size
is linked to the $\mathbf k$-point discretization grid.}
\label{fig:interactioncell}
\end{figure}

To better understand the approximations introduced by a finite
$\mathbf k$-point sampling, we will first derive the general
connection between the $\mathbf k$-point sampling and the range of 
nonlocality, and then discuss its relevance for the present case.
As an introductory remark, we note that in periodic systems
the two-point functions
$F=G$, $W$, or $\Sigma$ reflect the lattice periodicity
\begin{equation}
F(\mathbf r, \mathbf r') = 
F(\mathbf r + \mathbf R, \mathbf r' + \mathbf R)
\;,
\end{equation}
where $\mathbf R$ denotes a vector of the real-space lattice
defined by the unit-cell vectors $\mathbf a_1$, $\mathbf a_2$, 
and $\mathbf a_3$.
We can then introduce the representation
\begin{equation}
F_{\mathbf R}(\mathbf r, \mathbf r') := F(\mathbf r + \mathbf R, \mathbf r')
\;,
\end{equation}
where $\bf r$ and $\bf r'$ are restricted to the unit cell.
The corresponding reciprocal-space representation $F_{\mathbf k}$
(also denoted as mixed-space representation \cite{Rieger}) is obtained from
a Fourier transformation
\begin{equation}
F_{\mathbf k}(\mathbf r, \mathbf r') 
:= e^{-i\mathbf k\cdot (\mathbf r - \mathbf r')}\sum_{\mathbf R}
F_{\mathbf R}(\mathbf r, \mathbf r') \,e^{-i\mathbf k \cdot \mathbf R}
\;.
\end{equation}

%Formally, $\mathbf k$ is a continuous index over the
%Brillouin zone.
We denote the unit vectors of the reciprocal lattice 
by $\mathbf b_i, i \in \{1,2,3\}$. They are defined by 
those of the real-space unit cell via
$\mathbf a_i \mathbf b_j = 2\pi\delta_{ij}$.
The connection to the
$\mathbf R$ representation provides a real-space picture
of the $\mathbf k$-point discretization.
For this, we note that a regular, $\Gamma$-centered 
discretization grid ($N_1\times N_2 \times N_3$) defines
a lattice $\mathbf b_i/N_i$ in reciprocal space
that is associated with a real-space supercell $N_i \mathbf a_i$,
comprising $N_1 \times N_2 \times N_3$ unit cells
(cf. Fig.~\ref{fig:interactioncell}).
This supercell coincides with the interaction cell of the space-time
method \cite{gwstPRL}, or in other words the range of nonlocality  for
the two-point function $F=G$, $W$, and $\Sigma$. 
Discretizing $\mathbf k$ for the function $F_\mathbf k$
is therefore equivalent to imposing a translational symmetry
\begin{equation}
F_{\mathbf R}(\mathbf r, \mathbf r') = 
F_{\mathbf R + \mathbf S}(\mathbf r, \mathbf r')
\; ,
\label{eq:transsym}
\end{equation}
in real space,
where $\mathbf S$ denotes a lattice vector of the interaction-cell lattice
$(N_i \mathbf a_i)$.

From this connection we conclude that convergence in the 
$\mathbf k_\parallel$ sampling can only be achieved when the interaction
cell associated with the sampling is large enough to encompass the dominant
nonlocality in the relevant functions ($G$, $P$, $W$, and $\Sigma$).
This also holds true for $GW$ schemes that do not explicitly
rely on the real-space interaction cell, or which employ a
representation other than plane waves for the unit-cell coordinates 
($\mathbf r$, $\mathbf r'$). Let us now consider the behavior of the
different two-point functions.
In nonmetallic bulk systems,
the Green's function decays exponentially with increasing distance of
its arguments, typically over a few bond lengths \cite{Schindlmayr}. 
We see no reason to believe that repeated-slab systems
show a qualitatively different behavior in this respect.
However, the surface may exhibit a different band gap than the
bulk, or it may even be metallic. The decay properties of the Green's function
will then change accordingly \cite{Schindlmayr,IsmailBeigiArias}.
In particular, a reduction in the gap at the surface
implies a slower decay of $G$ at the surface compared to the bulk.
The decay properties of $G$ are directly transferred to $P$ 
[Eq.~(\ref{eq:defP})] and $\Sigma$ [Eq.~(\ref{eq:sig})].
The screened interaction $W$, on the other hand, exhibits
a very slow $1/r$ decay. 
However, this long-range limit $W^{\rm lr}$ is subtracted
from the full $W$ in the reciprocal-space singularity treatment
discussed previously. 
For the numerical convergence only the remainder $W^{\rm sr}$ is
relevant, i.e., the difference between $W$ and the model
function $W^{\rm lr}$ used in the singularity treatment.
$W^{\rm sr}$ is dominated by the variation of the
electron density and thus by the atomic structure, and should become
negligible when the typical length scale of these variations is exceeded.
In bulk systems, this corresponds to a few bond lengths.
However, the decisive structural variation in a repeated-slab system
is the slab itself (at the slab boundary the dielectric constant drops from
$\varepsilon_{\rm b}$ to 1). The thickness of the slab or the vacuum
therefore defines the length scale for the interaction-cell convergence
also parallel to the surface.

\begin{figure}
\center
\includegraphics[width=0.95\columnwidth]{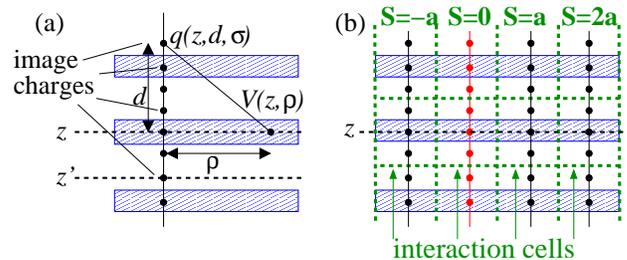}
\caption{a) Computation of the screened potential from the image charges.
b) Periodic repetition of the image charges to simulate the effect of the
$\mathbf k_\parallel$ discretization. $a$ denotes the
lateral extent of the interaction cell.}
\label{fig:imagecharges}
\end{figure}

To understand how the $\mathbf k_\parallel$ discretization 
modifies the screened interaction, we employ the classical theory
of dielectric screening. Similar to what was done above for
the macroscopic dielectric tensor, we approximate the slab by a
homogeneous dielectric with a sharp interface to the vacuum as sketched 
out in Fig.~\ref{fig:imagecharges}.
We then apply the method of image charges (cf. Appendix)
to compute the screening function.
When a charge is placed at point $z'$, 
the resulting image charges lie on a line through the original charge
perpendicular to the surface (cf. Fig.~\ref{fig:imagecharges}).
The effect of the $\mathbf k_\parallel$ discretization 
is simulated by imposing translational symmetry for
the image charges in the direction parallel to the surface
according to Eq.~(\ref{eq:transsym}). 
In other words, we explicitly consider a lattice of
perpendicularly aligned image charges [cf. Fig.~\ref{fig:imagecharges}(b)]
with the interaction cell's lattice constant of $N_\parallel$ times
the original cell's lattice constant.
To simplify the discussion we focus on one characteristic
aspect of the screened interaction: the potential $V^{\rm im}$
generated by the image charges at the position of the original charge,
i.e., at $z=z'$.
We obtain
\begin{equation}
V^{\rm im}(z) = \sum_{\mathbf S} 
\sum_{d,\sigma} \frac{q(z,d,\sigma)}{\sqrt{d^2 + |\mathbf S|^2}}
\label{eq:parerror_full}
,
\end{equation}
where the sum over $\mathbf S$ runs over the lattice vectors
of the interaction cell $N_{\parallel} \mathbf a_\parallel$.
$d$ and $\sigma$ enumerate the image charges that are computed
as described in the Appendix.
From this the long-range model function $1/(\varepsilon_\parallel r)$
and the contribution of the
original image charges at \mbox{$\mathbf S=\mathbf 0$} are subtracted.
Exploiting the sum
rule Eq.~(\ref{eq:sumrule}), we obtain
for the error introduced by the parallel discretization 
\begin{equation}
\Delta V^{\rm im}(z) = \sum_{\mathbf S \ne \mathbf 0} 
\sum_{d,\sigma} q(z,d,\sigma)\big(\frac{1}{\sqrt{d^2 + |\mathbf S|^2}}
      - \frac{1}{|\mathbf S|}\big)
\;.
\label{eq:modelparerr}
\end{equation}

\begin{figure}
\includegraphics[width=0.95\columnwidth,clip]{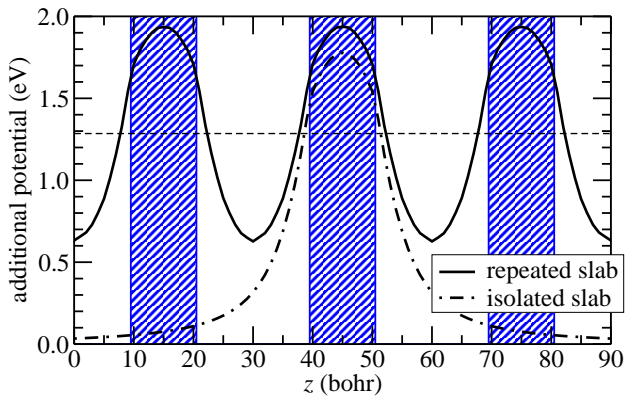}
\caption{(Color online) Additional potential $\Delta V^{\rm im}$ introduced
by the $\mathbf k_\parallel$ discretization
of the screened interaction (see text).
The dielectric model corresponds to a two-layer NaCl(100) slab in a 
$c$=30\,bohr 
unit cell
with a 3$\times$3 $\mathbf k_\parallel$ sampling.
The blue-shaded areas indicate the
position of the slabs; the horizontal dashed line marks the average potential.
The isolated-slab case is shown for comparison.}
\label{fig:parerror}
\end{figure}

In Fig.~\ref{fig:parerror} we show $\Delta V^{\rm im}(z)$ for the
dielectric model corresponding to the two-layer NaCl(100) slab
with a $3\times3$ $\mathbf k_\parallel$ sampling
 for both the isolated and the repeated-slab case.
The absolute error in the potential is very large.
We note that the average shift in the potential is implicitly corrected
for in the full $GW$ calculations by setting the 
$W_{\mathbf 0}(\mathbf 0, \mathbf 0)$ element to zero
after the subtraction of the model function. Therefore, only the
deviation from this average (dashed line in Fig.~\ref{fig:parerror})
contributes to the $\mathbf k$-point convergence,
which is considerably smaller but far from negligible.
We conclude that the convergence of a $GW$ calculation with
respect to the $\mathbf k_\parallel$ sampling is crucially influenced
by the discretization error of the screened interaction.

The variation of $\Delta V^{\rm im}$ from its average value
may be taken as an indicator for the discretization error in the
$GW$ self-energy. We investigated how this changes as a function
of slab and vacuum thickness within the dielectric model.
When the vacuum region is increased, the variation increases because
$\Delta V^{\rm im}$ quickly decays in the vacuum region
(cf. Fig.~\ref{fig:parerror}), thereby pulling down
the average over the computational cell.
The periodic repetition (small vacuum) thus
introduces a highly advantageous compensation effect:
While the absolute value of $\Delta V^{\rm im}$ is quite
large in the slab, the average is dominated by the contribution
of the slab, which then cancels out a large part of the total 
discretization error.
The physical origin behind the slab and vacuum dependence of the
$\mathbf k_\parallel$ convergence lies in the scale dependence of the
screening in such an inhomogeneous system.
For distances much smaller than the slab thickness, a bulk-like screening
is expected inside the slab. However, in cells
with large vacuum separations the components of the dielectric tensor may
deviate considerably from this bulk limit, as illustrated in
Sec.~\ref{sec:aniso}. Since the treatment for the long-ranged part
of the screened interaction assumes this behavior at \emph{all} length
scales, a portion of this difference in screening enters the numerical
treatment of the short-ranged part $W^{\rm sr}$ and affects the 
$\mathbf k_{||}$ convergence. If, on the other hand, the vacuum separation
is small, the
dielectric screening of the repeated-slab system is close to bulklike and
the magnitude of the discretization error is reduced correspondingly. We
find this anticipated behavior fully confirmed for the slab systems
considered in this work (Figure~\ref{fig:nacl-vac} will show this explicitly
for the two-layer NaCl(100) slab).

To estimate the influence of the $W$ discretization on the quasiparticle
energies computed in the $GW$ space-time method more precisely,
we have developed a fit
function to describe the dependence of the band energies
on the $\mathbf k$-point sampling $N_{\parallel}$.
For this purpose, we assume that the $G_0W_0$ corrections
show the same functional dependence on $N_{\parallel}$ as 
$\Delta V^{\rm im}$ in Eq.~(\ref{eq:modelparerr}).
Retaining a single term of the summation, we arrive at a three-parameter function:
\begin{equation}
\label{eq:fitfunction}
\epsilon_n(N_\parallel) = \epsilon_n(\infty) + 
\frac{Q_n}{N_{\parallel}} - \frac{Q_n}{\sqrt{D_n^2 + N_{\parallel}^2}}
\;.
\end{equation}
The parameters $Q_n$, $D_n$, and $\epsilon_n(\infty)$ are determined 
for each state $n$ by
fitting to the $GW$ data. We find that this relatively simple form
accurately describes the convergence with respect to $N_{\parallel}$
for all slab systems studied. Typical examples of the quality of the
fit are shown in Fig.~\ref{fig:kcurves}.
We employ the fitting procedure to estimate
the remaining error at finite $\mathbf k_\parallel$ sampling, or to
extrapolate the converged value $\epsilon_n(\infty)$ in cases where
the error at finite sampling is unacceptably large.

\subsection{Periodic repetition: slab-slab interaction}
\label{sec:vac}

\begin{figure}
\center
\includegraphics[width=0.95\columnwidth,clip]{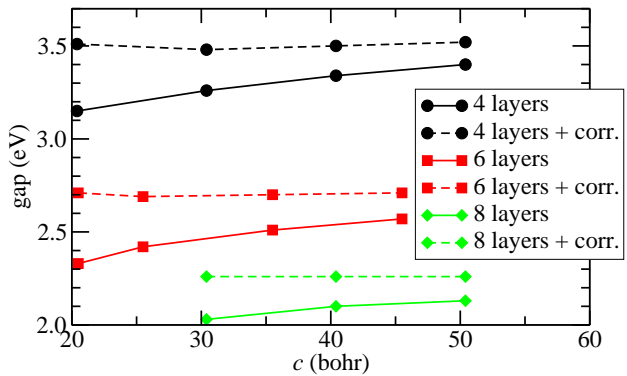}
\caption{Dependence of the band gap
of H-saturated Si(100) slabs with four, six, and eight layers
(two~layers $\approx$ 5\,bohr) on the total cell height $c$.
Solid lines: DFT+GW results. 
Dashed lines: DFT+GW results with finite-vacuum corrections derived
from the dielectric model.
The DFT gaps (not shown) are independent of the vacuum thickness.}
\label{fig:si-vac}
\end{figure}

Until now, we have focused the discussion on technical and numerical
aspects of $GW$ calculations for repeated-slab systems. We
now turn to the physical effects associated with the interslab 
polarization. 
For this, we computed the band structures for hydrogen-saturated Si(100)
slabs with four, six, and eight Si layers and varying amounts of vacuum. 
The $\mathbf k$-point convergence was tested for each system, based on the 
fitting procedure described above.
An $8\times8\times 1$ ($7\times7\times1$) sampling was found to be 
sufficient for the four-layer (six- and eight-layer) case.
The numerical results for the fundamental gap presented in 
Fig.~\ref{fig:si-vac} show a slow convergence with respect
to the vacuum size. The changes within the numerically accessible range
are noticeable (ranging between 0.05\,eV and 0.2\,eV per 10 bohr of vacuum).
More importantly, we estimate that vacuum regions of 80--100\,bohr would
be required to achieve absolute convergence within 0.05\,eV of the 
isolated-slab limit (which can be determined with the
correction scheme that will be described later).
However, only the gap between occupied and unoccupied states is affected.
The changes within the occupied or unoccupied
part of the spectrum are minor.
As we will show, this is a further effect of
the long-range screening that affects the occupied and unoccupied
bands as a whole: The presence of the neighboring slabs
gives rise to an additional contribution in the screened potential,
which decreases when the distance between the slabs is increased.

We investigated this within the dielectric model.
The model parameters are obtained consistently with the full $GW$ 
calculation from the dielectric tensor components $\varepsilon_\parallel$
and $\varepsilon_z$ and the total cell height $c$.
From Eqs.~(\ref{eq:eps_par}) and (\ref{eq:eps_z}), we find
\begin{eqnarray}
\varepsilon &=& \frac{\varepsilon_{\parallel} - 1}{1 - \varepsilon_z^{-1}} 
\label{eq:eps_EMT}\;,\\
s & = & c
\left[  \frac{1}{1 - \varepsilon_{\parallel}} 
      + \frac{1}{1 - \varepsilon_z^{-1}}\right]^{-1}
\label{eq:s_EMT}
\;.
\end{eqnarray}
To simulate the effect of the periodic repetition, we take into
account a finite number of slabs ($\approx$20) in our model calculation
and compare this with the limiting case of an isolated slab, which
corresponds to increasing the vacuum size to infinity.

The change in the self-energy that results
from the long-range screening effects can be estimated from the
static Coulomb-hole screened-exchange (COHSEX) approximation\cite{COHSEX}:
\begin{eqnarray}
\Sigma(\mathbf r,\mathbf r') &=& \Sigma_{\rm COH}(\mathbf r,\mathbf r')
+ \Sigma_{\rm SEX}(\mathbf r,\mathbf r')\;,\\
\Sigma_{\rm COH}(\mathbf r, \mathbf r') &=&
\frac 12 \delta(\mathbf r - \mathbf r')
\left[W(\mathbf r, \mathbf r') - v(\mathbf r, \mathbf r')\right]\;,\\
\Sigma_{\rm SEX}(\mathbf r,\mathbf r') &=&
-\sum_n^{\rm occ} \varphi_n(\mathbf r)\varphi^*_n(\mathbf r')W(\mathbf r,\mathbf r')
\;.
\end{eqnarray}
Here $W$ denotes the statically screened interaction and $v$ the bare Coulomb
interaction. To isolate the contribution arising from the neighboring
slabs we separate the
screened interaction into a bulklike part $W^{\rm bulk}$ and an additional 
contribution $W^{\rm pol}$ from long-range polarization effects.  
Since $\Sigma$ is a direct product of $W$ with the Green's function,
the self-energy reflects this separation:
\begin{equation}
\Sigma^{\rm pol}(\mathbf r,\mathbf r') = 
\left[ \frac 12 \delta(\mathbf r - \mathbf r') 
-\sum_n^{\rm occ} \varphi_n(\mathbf r)\varphi^*_n(\mathbf r')
\right]W^{\rm pol}(\mathbf r,\mathbf r')
\;.
\end{equation}
To simplify this expression further, we split off purely local
contributions from $W^{\rm pol}$ according to
\begin{eqnarray}
W^{\rm pol}(\mathbf r,\mathbf r') &=& \frac 12 \left[
W^{\rm pol}(\mathbf r,\mathbf r)
+ W^{\rm pol}(\mathbf r',\mathbf r')\right]
\label{eq:sigpolloc}
\\
&+& W^{\rm pol,nl}(\mathbf r, \mathbf r')
\end{eqnarray}
implicitly defining the purely nonlocal remainder $W^{\rm pol,nl}$.
We now argue that the self-energy $\Sigma^{\rm pol,nl}$
arising from this separation is negligible for long-range polarization effects.
In particular, we have by construction,
\begin{equation}
W^{\rm pol,nl}(\mathbf r, \mathbf r)=0\;.
\end{equation}
The corresponding COH contribution therefore vanishes.
We now turn to $\Sigma^{\rm pol,nl}_{\rm SEX}$.
Assuming that $W^{\rm pol}$ is generated by a
set of image charges at distance $d$, 
$W^{\rm pol,nl}(\mathbf r, \mathbf r')\approx0$ for 
$|\mathbf r-\mathbf r'|\ll d$. The magnitude of $d$ is determined
by the distance to the relevant dielectric interface. 
On the other hand, the second factor in the SEX term
is the Green's function at $t\rightarrow 0, t>0$. Its exponential
decay limits the magnitude of $|\mathbf r-\mathbf r'|$ to
a few interatomic distances \cite{Schindlmayr}.
Therefore, $\Sigma^{\rm pol,nl}_{\rm SEX}$ is expected to be small
for polarization effects taking place at a length scale of more than
a few bohr, and we can safely neglect its contribution.
For $\Sigma^{\rm pol,loc}$ arising from the local part
[Eq.~(\ref{eq:sigpolloc})],
the state summation reduces to a projection onto the occupied states,
and the expectation values of $\Sigma^{\rm pol,loc}$ become
\begin{equation}
\langle\varphi_n|\Sigma^{\rm pol,loc}|\varphi_n\rangle \pm\frac 12 \langle \varphi_n | W^{\rm pol}(\mathbf r, \mathbf r) | \varphi_n \rangle
\label{eq:deltaSig}
\; ,
\end{equation}
where the plus (minus) sign applies to unoccupied (occupied) states.
For the dielectric model, we now identify 
$W^{\rm bulk}$ with $1/(\varepsilon(z)|\mathbf r-\mathbf r'|)$.
$W^{\rm pol}(\mathbf r, \mathbf r)$ then reduces to 
the potential $V^{\rm im}(z)$ induced by a charge at its own 
position, the $\bf S=0$ term in Eq.~(\ref{eq:parerror_full}).
Equation~(\ref{eq:deltaSig}) corresponds to an adiabatic switching on of 
``image charge'' effects for the charged final state,
taking into account the sign convention of single-particle energies.
We note that Delerue \etal{}\cite{Delerue} have derived the same final
formula from an electrostatic model and used it to estimate
self-energy shifts in isolated nanoparticles.
% FOOTNOTE

\begin{figure}
\includegraphics[width=0.95\columnwidth,clip]{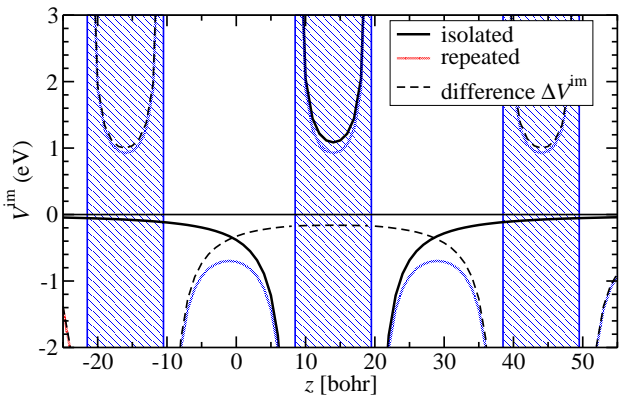}
\caption{Image potential for an isolated slab and a repeated-slab system
in comparison. The blue-shaded regions indicate the position of the
slabs. The model parameters correspond to a two-layer NaCl slab.}
\label{fig:iso-rep-comp}
\end{figure}

In Fig.~\ref{fig:iso-rep-comp}, we compare 
$V^{\rm im}(z)$ for an isolated slab with that of 
the repeated slabs. We find
for both cases a positive contribution inside the slab, and
a negative one outside, in agreement with Delerue \cite{Delerue}.
The divergence at the interface is
an artifact of the steplike dielectric profile
employed. In the repeated-slab case, however,
the potential is shifted downward because of the additional 
polarization in the neighboring slabs. It is precisely this polarization
of the neighboring
slabs that creates the undesired perturbation in the central slab and
is responsible for the observed
dependence of the band gap on the vacuum size.
The shift in the image potential (indicated by
the dashed line in Fig.~\ref{fig:iso-rep-comp})
is essentially constant over the slab and continuous across
the slab-vacuum interface.
The $\Sigma^{\rm pol}$ self-energy corrections according to
Eq.~(\ref{eq:deltaSig}) therefore do
not vary significantly between different states, 
giving rise to a scissor-like change in the gap
in agreement with the observations made for the full $G_0W_0$
calculation.
Moreover, this fortuitously simple situation facilitates
a quantitative comparison between the dielectric model and the dependence
of the band gap found in the $G_0W_0$ calculation.
To demonstrate this, we have extracted the magnitude of the 
finite-vacuum effect from the dielectric model and corrected the
numerical $G_0W_0$ data by these values for every vacuum thickness. 
The result has been included in
Fig.~\ref{fig:si-vac} as the dashed curves.
We find that this corrected data no
longer depends significantly on the vacuum size.
Using this correction scheme, it becomes possible to
determine the isolated-slab values for arbitrary vacuum sizes.

\begin{figure}
\center
\includegraphics[width=0.95\columnwidth,clip]{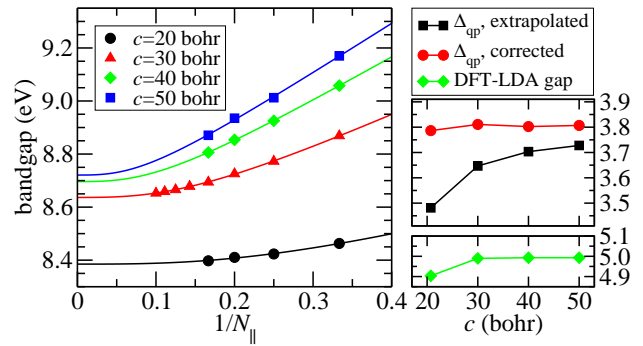}
\caption{Left: $\mathbf k_\parallel$ convergence for the gap
of a two-layer NaCl slab for different total cell heights $c$.
Right: DFT-LDA gap (bottom) and 
extrapolated quasiparticle correction $\Delta_{\rm qp}$ as well as its
value corrected for the finite vacuum (top)
as a function of $c$.}
\label{fig:nacl-vac}
\end{figure}

These \textit{a posteriori} corrections do not 
depend sensitively on the microscopic details of the material. We demonstrate
this for an ionic material, a two-layer NaCl(100) slab.
The dependence of the $\mathbf k_\parallel$ convergence behavior
on the slab/vacuum ratio (left side of Fig.~\ref{fig:nacl-vac})
agrees very well with
the predictions from the dielectric model: With increasing
vacuum size, the convergence becomes
more difficult.
For this reason, we found it necessary to extrapolate the
$\mathbf k_\parallel$
convergence for each vacuum thickness using Eq.~(\ref{eq:fitfunction}),
because
its magnitude becomes comparable to that of the finite-vacuum 
effect.
Using the extrapolated values, we find a very
good agreement for the finite-vacuum effect compared to
the dielectric model. This is demonstrated by the
quasiparticle correction $\Delta_{\rm qp}$ to the band gap 
shown at the right-hand side of Fig.~\ref{fig:nacl-vac}. Including the
finite-vacuum correction, $\Delta_{\rm qp}$ coincidentally converges even
faster with respect to the
vacuum size than the band gap of the underlying DFT-LDA 
calculation in this case.

\section{Discussion and Conclusions}
\label{sec:discussion}

In this work we have investigated the performance of the repeated-slab
approach in $G_0W_0$ calculations.
Using
the classic theory of dielectric screening we have found that
the relevant effects can be understood from simple dielectric slab 
models. 
We have demonstrated that long-range
polarization effects introduce several important differences in
slab systems compared to the corresponding bulk systems for
prototypical semiconductor (Si) and ionic (NaCl) materials.
The periodic repetition of the slabs introduces an additional
scissorlike change of the one-particle spectrum that can be
quantitatively corrected for \textit{a posteriori}. These
corrections are derived from dielectric slab models.

Since long-range effects are described by the small-$\mathbf k$
behavior in reciprocal space, this limit requires special
attention for slab calculations. The anisotropy
of the macroscopic screening at $\mathbf k\rightarrow \mathbf 0$
and the slow convergence with respect to the $\mathbf k_\parallel$
grid must be taken into account. 
If this is not done correctly,
it may not be possible to assess the
importance of the vacuum separation. However, a
careful treatment of these aspects does not pose 
principal problems and can be achieved with moderate computational
effort, in particular for small vacuum sizes. Using the
finite-vacuum correction scheme provides a highly efficient
technique to extract accurate values for the isolated-slab limit from
only a few calculations for small vacuum separations.

We will briefly discuss how far the issues raised in this work
for the $GW$ space-time method are relevant also for other 
computational schemes. The anisotropy of the dielectric tensor
is specific to the repeated-slab approach when no Coulomb-truncation
techniques are employed. We have demonstrated recently
\cite{myCPC} how the anisotropy can be treated to arbitrary precision in
plane-wave approaches by expanding the angular dependence in spherical
harmonics. In practice, already a maximum angular momentum of 2 is
sufficient to capture most of the effect \cite{myCPC}. We expect
that a similar strategy is applicable for other basis sets, too.

The slow convergence with respect to the $\mathbf k_\parallel$ sampling
is rooted in the reciprocal-space computation
of the screened interaction. To our knowledge, all $GW$ implementations
with periodic boundary conditions rely on this strategy.
However, studying the convergence with respect to this parameter
is computationally very efficient in the space-time approach since
it scales linearly in the number of $\mathbf k$ points, compared
to a typically quadratic scaling in convolution approaches.
The discretization error arises from the part that is treated numerically,
i.e., the deviation of the screened interaction from the model
function used for the $1/k^2$ singularity treatment.
Alternative integration schemes,
such as the improved integration scheme by Pulci \etal\cite{Pulci}
or the offset $\Gamma$-point method \cite{Lebegue}, will probably not 
improve this significantly, because they aim at better integrating 
the $1/(\mathbf k^{\rm T}\mathbf L\mathbf k)$ model function,
not the deviations from it.
The $\mathbf k_\parallel$ convergence is a critical issue
also for isolated slabs when treated by 
Coulomb-truncation techniques.
In fact, it becomes even more dramatic in these cases because the
long-range tail then equals the unscreened $1/r$ Coulomb interaction
and the small-vacuum compensation effect is excluded.
Indeed, Coulomb-truncation techniques
have been found to require drastically enlarged $\mathbf k$-point 
grids in the nontruncated directions \cite{LouieNanoWire,Rozzi,Sohrab}.
We note that the anisotropy and the $\mathbf k$-point convergence
are relevant also for condensed slablike systems, such as multilayers
or quantum-well superlattices, but their magnitude depends critically
on the dielectric constants of the materials involved.

The dependence of the gap on the vacuum thickness,
on the other hand, is a physical property of repeated-slab systems.
Since not only the closest neighboring slabs contribute to this
effect, its magnitude reduces with increasing slab thickness 
\cite{remark:s2vrule}.
It is reproduced with the model calculation and is thus
independent of the computational scheme used for the $G_0W_0$ calculations.
Any implementation that does not find this trend must employ
additional approximations that suppress this behavior. 
Whether such an implementation then provides the isolated-slab
limit depends on the actual approximations and cannot be foreseen in
general.

For future $GW$ calculations in repeated-slab systems, we
recommend the following procedure:
\begin{enumerate}
\item Use an anisotropic treatment of the screened Coulomb singularity 
for systematic $\mathbf k$-point convergence studies.
\item Check the $\mathbf k$-point sampling parallel to the surface
separately for each slab and vacuum thickness.
If necessary, extrapolate using a physically justified function.
\item Correct band energies/gaps for the finite-vacuum thickness
according to Eq.~(\ref{eq:deltaSig}) and the Appendix.%~\ref{app:imgch}.
Equations~(\ref{eq:eps_EMT}) and (\ref{eq:s_EMT}) provide model parameters
consistent with the actual dielectric tensor.
\end{enumerate}

For surface calculations, it should be kept in mind that the
surface electronic structure of a
slab differs from that of a semi-infinite solid
due to confinement of the electronic states.
This aspect is already present at the level of DFT and is directly
transferred to the $GW$ band structure.
Quantum confinement may for instance split a surface
resonance into a series of quantized states. In addition
states localized at the two slab surfaces may couple through
the slab at insufficient slab thicknesses. 
$GW$ corrections might become important if they shift the
energies of surface states or resonances relative to the bulk,
as this alters the decay behavior.

For the surface $GW$ self-energy itself, the role of the slab approximation
has, to our knowledge, not been studied. An in-depth analysis of this
question goes beyond the scope of the present paper, but
we will briefly discuss the relevant aspects for $G$ and $W$.
For semiconductors the exponential
spatial decay of the Green's function leads to a rapidly
decreasing influence of the second surface with increasing slab thickness. 
However, if this decrease is fast enough to be unimportant at the 
commonly employed slab thicknesses of only a few atomic layers
cannot be guaranteed {\it a priori}. 
For the screened interaction, on the other hand, the influence
of the dielectric discontinuity decays only slowly.
Using simple electrostatic models similar to those detailed in 
Sec.~\ref{sec:park} (see also Delerue \etal{} \cite{Delerue}{}), 
we find that the presence of the additional surface at
a distance $s$ gives rise to an image-charge-like contribution
$\approx \frac{\varepsilon_{\rm b} - 1}{2\varepsilon_{\rm b}
(\varepsilon_{\rm b}+1) s}$. Unlike for the finite-vacuum
effect (Sec.~\ref{sec:vac}), however, quantitative corrections
for this \emph{finite-slab screening effect} depend strongly on
the spatial extent of the electronic states. 

In summary, we find that the repeated-slab approach provides
a computationally efficient way to calculate electron and hole
energies in the $G_0W_0$ approximation for two-dimensional systems
like surfaces and thin films.
Three manifestations of long-range screening effects in two dimensions ---
the anisotropy of the macroscopic screening,
the $\mathbf k_\parallel$ sampling and
the periodic repetition of the slabs ---
have been identified and analyzed in terms of the classic theory of
dielectric screening.
Their effect on  $G_0W_0$ calculations in the repeated-slab
approach has been demonstrated for thin Si and NaCl films.
Robust and efficient correction schemes have been developed.
With these, the isolated-slab limit can be easily obtained from
repeated-slab calculations.

\section*{Acknowledgments}
We thank Lucia Reining, Fabien Bruneval, Francesco Sottile, Carlo Rozzi, 
Hardy Gross, Angel Rubio, and Christoph Friedrich
for fruitful discussions. This work was funded in part by the EU
through the Nanophase Research Training Network (Contract No. 
HPRN-CT-2000-00167) and the Nanoquanta Network of Excellence 
(Contract No. NMP-4-CT-2004-500198).
P.E. acknowledges the Deutscher Akademischer Aus\-tausch\-dienst
for financial support.

\appendix
\section*{Computation of image charges in multilayer systems}
\label{app:imgch}

In this section we derive a practical scheme to compute the
screened interaction
in a dielectric multilayer system using image charges (cf. Fig.~\ref{fig:imagecharges}).
The screened interaction $W(\mathbf r, \mathbf r')$ is obtained
as the additional potential $V(\mathbf r)$ induced when a unit charge 
is placed at $\mathbf r'$.
$V(\mathbf r)$ is constructed with a series of image charges
vertically aligned above and below the original position.

Let us first consider the textbook situation of two
semi-infinite dielectric media $\Omega_1$ and $\Omega_2$ 
with dielectric constants $\varepsilon_1$ and $\varepsilon_2$ \cite{Jackson}.
For a charge $q$ at $\mathbf r'$ in $\Omega_1$,
the potential $V(\bf r)$ is given by
\begin{eqnarray}
\mathbf r \in \Omega_1\,\textnormal{:}\quad
V(\mathbf r) &= & \frac{1}{\varepsilon_1}\left(\frac{q}{|\mathbf r - \mathbf r'|} 
+ \frac{q''}{|\mathbf r - \mathbf r''|}\right)
\; ,
\\
\mathbf r \in \Omega_2\,\textnormal{:}\quad
V(\mathbf r) &=& \frac{1}{\varepsilon_2}\,\frac{q'}{|\mathbf r - \mathbf r'|} 
\;,
\end{eqnarray}
where $q'$ and $q''$ are image charges.
$\mathbf r''$ is obtained by reflecting $\mathbf r'$ at the interface,
and we will therefore denote $q''$ as ``reflected charge,'' whereas
the effect of $q$ is propagated into $\Omega_2$ by the 
``propagated charge'' $q'$.
They are determined from the continuity equations of the electric 
field and the electric displacement at the interface, yielding
\begin{equation}
q' = \frac{2 \varepsilon_2}{\varepsilon_1 + \varepsilon_2} \,q
%\textnormal{\qquad(\theequation)}\refstepcounter{equation}
\hspace{0.5cm}\textnormal{and}
\hspace{0.5cm}
q''= \frac{\varepsilon_1 - \varepsilon_2}{\varepsilon_1 + \varepsilon_2}\,q
\label{eq:qpr}
\label{eq:qrf}
\;.
\end{equation}

In order to develop a computational scheme for a multi-layer system, a proper
book-keeping is crucial to track the reflection and propagation of the various image charges.
For this purpose, we divide the system into individual layers.
Each layer has the same thickness $L$ and a layer-specific dielectric
constant $\varepsilon(z)$ \cite{eps_profile}.
For the model calculations in this work, we use $L$=1 bohr.
We will restrict the notation to the $z$ coordinate for 
the position of the charges and layers. The parallel coordinate
$\rho$ becomes only relevant for the computation of the potential;
see Fig.~\ref{fig:imagecharges}(a).
The origin of the coordinate system is chosen such that the layers are
centered around integers $L$, and the interfaces are at half integers.

For reasons that will become clear below, we denote an image charge
that contributes to the potential in layer $z$ and is located at 
$z + \sigma d$ by $q(z,d,\sigma)$, where
$d \ge 0$ is the distance from the layer and $\sigma=\pm1$.
To show that the image charges can be determined iteratively
we will now derive the iteration for $d\rightarrow d+L$.
Consider a charge $q(z,d,\sigma)$ relevant for the potential in 
layer $z$. Due to the interface at 
$z-\frac12\sigma L$ two additional image charges appear.
The reflected image charge, located at $z -\sigma(d+L)$,
describes the potential in layer $z$ and is given by [cf. (Eq.~\ref{eq:qrf})]
\begin{equation}
q^{\rm rf}(z,d+L,-\sigma) 
= \frac{\varepsilon(z)-\varepsilon(z-\sigma)}
       {\varepsilon(z)+\varepsilon(z-\sigma)}\, q(z,d,\sigma)
\; .
\end{equation}
The propagated image charge remains at the position $z+\sigma d$
and describes the potential in layer $z - \sigma L$.
Using our book-keeping notation and Eq.~(\ref{eq:qpr}),
it can be written as
\begin{equation}
q^{\rm pr}(z-\sigma,d+L,\sigma)
= \frac{2 \varepsilon(z-\sigma)}
       {\varepsilon(z)+\varepsilon(z-\sigma)}\, q(z,d,\sigma)
\;.
\end{equation}
Obviously, the distance parameter is increased by $L$ for each interface
taken into account.
The image charges for the distance $d+L$ can thus be computed iteratively 
from those at distance $d$ and will in general combine
a reflected and
a propagated contribution $q=q^{\rm rf}+q^{\rm pr}$.
The iterations are started by setting
\begin{equation}
q(z',0,\pm1) = 1 \; .
\end{equation}
We restrict $z'$ to integer $L$, i.e., the original charge is
placed at the center of a layer, thereby avoiding the divergence
of the image potential at the dielectric discontinuities between adjacent
layers.
In practice, the iterations are stopped at some $d_{\rm max}$,
which thereby becomes a convergence parameter.
In addition, we truncate the number of layers considered explicitly
and neglect image charges that fall outside.
This truncation becomes a second convergence parameter.
The convergence for both parameters was tested by doubling the parameter
until the changes became negligible.

Summing the Coulomb potential of all the image charges
relevant for this layer, one obtains
\begin{eqnarray}
V(z,\rho) &=& \frac{1}{\varepsilon(z)}
\Big(\frac{\delta_{zz'}}{\rho}
%\nonumber\\ &&
+\sum_{d=L}^{d_{\rm max}}\sum_{\sigma=\pm1} \frac{q(z,d,\sigma)}
{\sqrt{d^2+\rho^2}}\Big)
\label{eq:screenedPotential}
\;.
\end{eqnarray}
The first term in Eq.~(\ref{eq:screenedPotential}) describes the 
potential of the original unit charge for $z=z'$ and corresponds
to that of a bulk material with the local dielectric 
constant $\varepsilon(z)$. The
sum over image charges $q(z,d,\sigma)$ then introduces the 
long-range polarization effects due to the variations in the
dielectric constant.

As a final remark, we mention a useful connection to the 
macroscopic anisotropy.
The long-range limit in the direction parallel to the surface 
$1/(\varepsilon_{\parallel}\,\rho)$ is obtained for $\rho\gg d$
by neglecting the perpendicular distance $d$ in the denominator of 
Eq.~(\ref{eq:screenedPotential}). From this, we obtain the sum rule
\begin{equation}
\frac{1}{\varepsilon(z)}
\left(\delta_{zz'} + \sum_{d,\sigma} q(z,d,\sigma)\right)
   = \frac{1}{\varepsilon_{\parallel}}
\label{eq:sumrule}
\;.
\end{equation}

%\bibliography{./slabgw}

\begin{thebibliography}{46}
\expandafter\ifx\csname natexlab\endcsname\relax\def\natexlab#1{#1}\fi
\expandafter\ifx\csname bibnamefont\endcsname\relax
  \def\bibnamefont#1{#1}\fi
\expandafter\ifx\csname bibfnamefont\endcsname\relax
  \def\bibfnamefont#1{#1}\fi
\expandafter\ifx\csname citenamefont\endcsname\relax
  \def\citenamefont#1{#1}\fi
\expandafter\ifx\csname url\endcsname\relax
  \def\url#1{\texttt{#1}}\fi
\expandafter\ifx\csname urlprefix\endcsname\relax\def\urlprefix{URL }\fi
\providecommand{\bibinfo}[2]{#2}
\providecommand{\eprint}[2][]{\url{#2}}

\bibitem[{\citenamefont{Hedin}(1965)}]{Hedin}
\bibinfo{author}{\bibfnamefont{L.}~\bibnamefont{Hedin}},
  \bibinfo{journal}{Phys.\ Rev.} \textbf{\bibinfo{volume}{139}},
  \bibinfo{pages}{A796} (\bibinfo{year}{1965}).

\bibitem[{\citenamefont{Aulbur et~al.}(2000)\citenamefont{Aulbur, J\"onsson,
  and Wilkins}}]{Aulbur}
\bibinfo{author}{\bibfnamefont{W.~G.} \bibnamefont{Aulbur}},
  \bibinfo{author}{\bibfnamefont{L.}~\bibnamefont{J\"onsson}},
  \bibnamefont{and} \bibinfo{author}{\bibfnamefont{J.~W.}
  \bibnamefont{Wilkins}}, \bibinfo{journal}{Solid State Phys.: Adv. Res. and
  Appl.} \textbf{\bibinfo{volume}{54}}, \bibinfo{pages}{1}
  (\bibinfo{year}{2000}).

\bibitem[{\citenamefont{Rinke et~al.}(2005)\citenamefont{Rinke, Qteish,
  Neugebauer, Freysoldt, and Scheffler}}]{Rinke}
\bibinfo{author}{\bibfnamefont{P.}~\bibnamefont{Rinke}},
  \bibinfo{author}{\bibfnamefont{A.}~\bibnamefont{Qteish}},
  \bibinfo{author}{\bibfnamefont{J.}~\bibnamefont{Neugebauer}},
  \bibinfo{author}{\bibfnamefont{C.}~\bibnamefont{Freysoldt}},
  \bibnamefont{and}
  \bibinfo{author}{\bibfnamefont{M.}~\bibnamefont{Scheffler}},
  \bibinfo{journal}{New J. Phys.} \textbf{\bibinfo{volume}{7}},
  \bibinfo{pages}{126} (\bibinfo{year}{2005}).

\bibitem[{\citenamefont{Garc\'ia-Gonz\'alez and Godby}(2001)}]{GWsurfaces}
\bibinfo{author}{\bibfnamefont{P.}~\bibnamefont{Garc\'ia-Gonz\'alez}}
  \bibnamefont{and} \bibinfo{author}{\bibfnamefont{R.~W.} \bibnamefont{Godby}},
  \bibinfo{journal}{Comput. Phys. Commun.} \textbf{\bibinfo{volume}{137}},
  \bibinfo{pages}{108} (\bibinfo{year}{2001}).

\bibitem[{\citenamefont{Rohlfing et~al.}(2003)\citenamefont{Rohlfing, Wang,
  Kr\"uger, and Pollmann}}]{InsulatorImageStates}
\bibinfo{author}{\bibfnamefont{M.}~\bibnamefont{Rohlfing}},
  \bibinfo{author}{\bibfnamefont{N.-P.} \bibnamefont{Wang}},
  \bibinfo{author}{\bibfnamefont{P.}~\bibnamefont{Kr\"uger}}, \bibnamefont{and}
  \bibinfo{author}{\bibfnamefont{J.}~\bibnamefont{Pollmann}},
  \bibinfo{journal}{Phys. Rev. Lett.} \textbf{\bibinfo{volume}{91}},
  \bibinfo{pages}{256802} (\bibinfo{year}{2003}).

\bibitem[{\citenamefont{Neaton et~al.}(2006)\citenamefont{Neaton, Hybertsen,
  and Louie}}]{benzeneGraphite}
\bibinfo{author}{\bibfnamefont{J.~B.} \bibnamefont{Neaton}},
  \bibinfo{author}{\bibfnamefont{M.~S.} \bibnamefont{Hybertsen}},
  \bibnamefont{and} \bibinfo{author}{\bibfnamefont{S.~G.} \bibnamefont{Louie}},
  \bibinfo{journal}{Phys. Rev. Lett.} \textbf{\bibinfo{volume}{97}},
  \bibinfo{pages}{216405} (\bibinfo{year}{2006}).

\bibitem[{\citenamefont{Hedstr\"om et~al.}(2006)\citenamefont{Hedstr\"om,
  Schindlmayr, Schwarz, and Scheff\-ler}}]{GWforDefectsPRL}
\bibinfo{author}{\bibfnamefont{M.}~\bibnamefont{Hedstr\"om}},
  \bibinfo{author}{\bibfnamefont{A.}~\bibnamefont{Schindlmayr}},
  \bibinfo{author}{\bibfnamefont{G.}~\bibnamefont{Schwarz}}, \bibnamefont{and}
  \bibinfo{author}{\bibfnamefont{M.}~\bibnamefont{Scheff\-ler}},
  \bibinfo{journal}{Phys. Rev. Lett.} \textbf{\bibinfo{volume}{97}},
  \bibinfo{pages}{226401} (\bibinfo{year}{2006}).

\bibitem[{\citenamefont{Hybertsen and Louie}(1986)}]{HybertsenLouie}
\bibinfo{author}{\bibfnamefont{M.~S.} \bibnamefont{Hybertsen}}
  \bibnamefont{and} \bibinfo{author}{\bibfnamefont{S.~G.} \bibnamefont{Louie}},
  \bibinfo{journal}{Phys.\ Rev.\ B} \textbf{\bibinfo{volume}{34}},
  \bibinfo{pages}{5390} (\bibinfo{year}{1986}).

\bibitem[{\citenamefont{Rojas et~al.}(1995)\citenamefont{Rojas, Godby, and
  Needs}}]{gwstPRL}
\bibinfo{author}{\bibfnamefont{H.~N.} \bibnamefont{Rojas}},
  \bibinfo{author}{\bibfnamefont{R.~W.} \bibnamefont{Godby}}, \bibnamefont{and}
  \bibinfo{author}{\bibfnamefont{R.~J.} \bibnamefont{Needs}},
  \bibinfo{journal}{Phys.\ Rev.\ Lett.} \textbf{\bibinfo{volume}{74}},
  \bibinfo{pages}{1827} (\bibinfo{year}{1995}).

\bibitem[{\citenamefont{Leb\`egue et~al.}(2003)\citenamefont{Leb\`egue, Arnaud,
  Alouani, and Bloechl}}]{Lebegue}
\bibinfo{author}{\bibfnamefont{S.}~\bibnamefont{Leb\`egue}},
  \bibinfo{author}{\bibfnamefont{B.}~\bibnamefont{Arnaud}},
  \bibinfo{author}{\bibfnamefont{M.}~\bibnamefont{Alouani}}, \bibnamefont{and}
  \bibinfo{author}{\bibfnamefont{P.~E.} \bibnamefont{Bloechl}},
  \bibinfo{journal}{Phys.\ Rev.\ B} \textbf{\bibinfo{volume}{67}},
  \bibinfo{pages}{155208} (\bibinfo{year}{2003}).

\bibitem[{\citenamefont{Kotani and van\ Schilfgaarde}(2002)}]{Kotani}
\bibinfo{author}{\bibfnamefont{T.}~\bibnamefont{Kotani}} \bibnamefont{and}
  \bibinfo{author}{\bibfnamefont{M.}~\bibnamefont{van\ Schilfgaarde}},
  \bibinfo{journal}{Solid State Commun.} \textbf{\bibinfo{volume}{121}},
  \bibinfo{pages}{461} (\bibinfo{year}{2002}).

\bibitem[{\citenamefont{Rohlfing et~al.}(1995)\citenamefont{Rohlfing, Kr\"uger,
  and Pollmann}}]{Rohlfing}
\bibinfo{author}{\bibfnamefont{M.}~\bibnamefont{Rohlfing}},
  \bibinfo{author}{\bibfnamefont{P.}~\bibnamefont{Kr\"uger}}, \bibnamefont{and}
  \bibinfo{author}{\bibfnamefont{J.}~\bibnamefont{Pollmann}},
  \bibinfo{journal}{Phys.\ Rev.\ B} \textbf{\bibinfo{volume}{52}},
  \bibinfo{pages}{1905} (\bibinfo{year}{1995}).

\bibitem[{\citenamefont{Hybertsen and Louie}(1988)}]{HybertsenLouieSurfaces}
\bibinfo{author}{\bibfnamefont{M.~S.} \bibnamefont{Hybertsen}}
  \bibnamefont{and} \bibinfo{author}{\bibfnamefont{S.~G.} \bibnamefont{Louie}},
  \bibinfo{journal}{Phys. Rev. B} \textbf{\bibinfo{volume}{38}},
  \bibinfo{pages}{4033} (\bibinfo{year}{1988}).

\bibitem[{\citenamefont{Neugebauer and Scheffler}(1992)}]{DFTdipoleCorrection}
\bibinfo{author}{\bibfnamefont{J.}~\bibnamefont{Neugebauer}} \bibnamefont{and}
  \bibinfo{author}{\bibfnamefont{M.}~\bibnamefont{Scheffler}},
  \bibinfo{journal}{Phys. Rev. B} \textbf{\bibinfo{volume}{46}},
  \bibinfo{pages}{16067} (\bibinfo{year}{1992}).

\bibitem[{\citenamefont{Onida et~al.}(1995)\citenamefont{Onida, Reining, Godby,
  Del~Sole, and Andreoni}}]{Na4}
\bibinfo{author}{\bibfnamefont{G.}~\bibnamefont{Onida}},
  \bibinfo{author}{\bibfnamefont{L.}~\bibnamefont{Reining}},
  \bibinfo{author}{\bibfnamefont{R.~W.} \bibnamefont{Godby}},
  \bibinfo{author}{\bibfnamefont{R.}~\bibnamefont{Del~Sole}}, \bibnamefont{and}
  \bibinfo{author}{\bibfnamefont{W.}~\bibnamefont{Andreoni}},
  \bibinfo{journal}{Phys. Rev. Lett.} \textbf{\bibinfo{volume}{75}},
  \bibinfo{pages}{818} (\bibinfo{year}{1995}).

\bibitem[{\citenamefont{Spataru et~al.}(2004)\citenamefont{Spataru,
  Ismail-Beigi, Benedict, and Louie}}]{LouieNanoWire}
\bibinfo{author}{\bibfnamefont{C.~D.} \bibnamefont{Spataru}},
  \bibinfo{author}{\bibfnamefont{S.}~\bibnamefont{Ismail-Beigi}},
  \bibinfo{author}{\bibfnamefont{L.~X.} \bibnamefont{Benedict}},
  \bibnamefont{and} \bibinfo{author}{\bibfnamefont{S.~G.} \bibnamefont{Louie}},
  \bibinfo{journal}{Appl. Phys. A: Mater. Sci. Process.}
  \textbf{\bibinfo{volume}{78}}, \bibinfo{pages}{1129} (\bibinfo{year}{2004}).

\bibitem[{\citenamefont{Rozzi et~al.}(2006)\citenamefont{Rozzi, Varsano,
  Marini, Gross, and Rubio}}]{Rozzi}
\bibinfo{author}{\bibfnamefont{C.~A.} \bibnamefont{Rozzi}},
  \bibinfo{author}{\bibfnamefont{D.}~\bibnamefont{Varsano}},
  \bibinfo{author}{\bibfnamefont{A.}~\bibnamefont{Marini}},
  \bibinfo{author}{\bibfnamefont{E.~K.~U.} \bibnamefont{Gross}},
  \bibnamefont{and} \bibinfo{author}{\bibfnamefont{A.}~\bibnamefont{Rubio}},
  \bibinfo{journal}{Phys. Rev. B} \textbf{\bibinfo{volume}{73}},
  \bibinfo{pages}{205119} (\bibinfo{year}{2006}).

\bibitem[{\citenamefont{Ismail-Beigi}(2006)}]{Sohrab}
\bibinfo{author}{\bibfnamefont{S.}~\bibnamefont{Ismail-Beigi}},
  \bibinfo{journal}{Phys. Rev. B} \textbf{\bibinfo{volume}{73}},
  \bibinfo{pages}{233103} (\bibinfo{year}{2006}).

\bibitem[{\citenamefont{Fratesi et~al.}(2003)\citenamefont{Fratesi, Brivio,
  Rinke, and Godby}}]{Fratesi1}
\bibinfo{author}{\bibfnamefont{G.}~\bibnamefont{Fratesi}},
  \bibinfo{author}{\bibfnamefont{G.~P.} \bibnamefont{Brivio}},
  \bibinfo{author}{\bibfnamefont{P.}~\bibnamefont{Rinke}}, \bibnamefont{and}
  \bibinfo{author}{\bibfnamefont{R.~W.} \bibnamefont{Godby}},
  \bibinfo{journal}{Phys.\ Rev.\ B} \textbf{\bibinfo{volume}{68}},
  \bibinfo{pages}{195404} (\bibinfo{year}{2003}).

\bibitem[{\citenamefont{Fratesi et~al.}(2004)\citenamefont{Fratesi, Brivio, and
  Molinari}}]{Fratesi2}
\bibinfo{author}{\bibfnamefont{G.}~\bibnamefont{Fratesi}},
  \bibinfo{author}{\bibfnamefont{G.~P.} \bibnamefont{Brivio}},
  \bibnamefont{and} \bibinfo{author}{\bibfnamefont{L.~G.}
  \bibnamefont{Molinari}}, \bibinfo{journal}{Phys.\ Rev.\ B}
  \textbf{\bibinfo{volume}{69}}, \bibinfo{pages}{245113}
  (\bibinfo{year}{2004}).

\bibitem[{\citenamefont{Tiago and Chelikowsky}(2005)}]{Tiago}
\bibinfo{author}{\bibfnamefont{M.~L.} \bibnamefont{Tiago}} \bibnamefont{and}
  \bibinfo{author}{\bibfnamefont{J.~R.} \bibnamefont{Chelikowsky}},
  \bibinfo{journal}{Solid State Commun.} \textbf{\bibinfo{volume}{136}},
  \bibinfo{pages}{333} (\bibinfo{year}{2005}).

\bibitem[{\citenamefont{Tiago and Chelikowsky}(2006)}]{Tiago2}
\bibinfo{author}{\bibfnamefont{M.~L.} \bibnamefont{Tiago}} \bibnamefont{and}
  \bibinfo{author}{\bibfnamefont{J.~R.} \bibnamefont{Chelikowsky}},
  \bibinfo{journal}{Phys. Rev. B} \textbf{\bibinfo{volume}{73}},
  \bibinfo{pages}{205334} (\bibinfo{year}{2006}).

\bibitem[{\citenamefont{Friedrich et~al.}(2006)\citenamefont{Friedrich,
  Schindlmayr, Bl\"ugel, and Kotani}}]{FriedrichLAPW}
\bibinfo{author}{\bibfnamefont{C.}~\bibnamefont{Friedrich}},
  \bibinfo{author}{\bibfnamefont{A.}~\bibnamefont{Schindlmayr}},
  \bibinfo{author}{\bibfnamefont{S.}~\bibnamefont{Bl\"ugel}}, \bibnamefont{and}
  \bibinfo{author}{\bibfnamefont{T.}~\bibnamefont{Kotani}},
  \bibinfo{journal}{Phys. Rev. B} \textbf{\bibinfo{volume}{74}},
  \bibinfo{pages}{045104} (\bibinfo{year}{2006}).

\bibitem[{\citenamefont{Bruneval et~al.}(2006)\citenamefont{Bruneval, Vast, and
  Reining}}]{Bruneval}
\bibinfo{author}{\bibfnamefont{F.}~\bibnamefont{Bruneval}},
  \bibinfo{author}{\bibfnamefont{N.}~\bibnamefont{Vast}}, \bibnamefont{and}
  \bibinfo{author}{\bibfnamefont{L.}~\bibnamefont{Reining}},
  \bibinfo{journal}{Phys. Rev. B} \textbf{\bibinfo{volume}{74}},
  \bibinfo{pages}{045102} (\bibinfo{year}{2006}).

\bibitem[{\citenamefont{Shishkin and Kresse}(2007)}]{KresseGW}
\bibinfo{author}{\bibfnamefont{M.}~\bibnamefont{Shishkin}} \bibnamefont{and}
  \bibinfo{author}{\bibfnamefont{G.}~\bibnamefont{Kresse}},
  \bibinfo{journal}{Phys. Rev. B} \textbf{\bibinfo{volume}{75}},
  \bibinfo{pages}{235102} (\bibinfo{year}{2007}).

\bibitem[{\citenamefont{Faleev et~al.}(2004)\citenamefont{Faleev, van
  Schilfgaarde, and Kotani}}]{SchilfgaardeSCGW}
\bibinfo{author}{\bibfnamefont{S.~V.} \bibnamefont{Faleev}},
  \bibinfo{author}{\bibfnamefont{M.}~\bibnamefont{van Schilfgaarde}},
  \bibnamefont{and} \bibinfo{author}{\bibfnamefont{T.}~\bibnamefont{Kotani}},
  \bibinfo{journal}{Phys. Rev. Lett.} \textbf{\bibinfo{volume}{93}},
  \bibinfo{pages}{126406} (\bibinfo{year}{2004}).

\bibitem[{\citenamefont{Delaney et~al.}(2004)\citenamefont{Delaney,
  Garc\'{\i}a-Gonz\'alez, Rubio, Rinke, and Godby}}]{Delaney}
\bibinfo{author}{\bibfnamefont{K.}~\bibnamefont{Delaney}},
  \bibinfo{author}{\bibfnamefont{P.}~\bibnamefont{Garc\'{\i}a-Gonz\'alez}},
  \bibinfo{author}{\bibfnamefont{A.}~\bibnamefont{Rubio}},
  \bibinfo{author}{\bibfnamefont{P.}~\bibnamefont{Rinke}}, \bibnamefont{and}
  \bibinfo{author}{\bibfnamefont{R.~W.} \bibnamefont{Godby}},
  \bibinfo{journal}{Phys.\ Rev.\ Lett.} \textbf{\bibinfo{volume}{93}},
  \bibinfo{pages}{249701} (\bibinfo{year}{2004}).

\bibitem[{\citenamefont{Biermann et~al.}(2003)\citenamefont{Biermann,
  Aryasetiawan, and Georges}}]{Biermann}
\bibinfo{author}{\bibfnamefont{S.}~\bibnamefont{Biermann}},
  \bibinfo{author}{\bibfnamefont{F.}~\bibnamefont{Aryasetiawan}},
  \bibnamefont{and} \bibinfo{author}{\bibfnamefont{A.}~\bibnamefont{Georges}},
  \bibinfo{journal}{Phys.\ Rev.\ Lett.} \textbf{\bibinfo{volume}{90}},
  \bibinfo{pages}{086402} (\bibinfo{year}{2003}).

\bibitem[{\citenamefont{Sun and Kotliar}(2004)}]{Sun}
\bibinfo{author}{\bibfnamefont{P.}~\bibnamefont{Sun}} \bibnamefont{and}
  \bibinfo{author}{\bibfnamefont{G.}~\bibnamefont{Kotliar}},
  \bibinfo{journal}{Phys.\ Rev.\ Lett.} \textbf{\bibinfo{volume}{92}},
  \bibinfo{pages}{196402} (\bibinfo{year}{2004}).

\bibitem[{\citenamefont{Onida et~al.}(2002)\citenamefont{Onida, Reining, and
  Rubio}}]{Onida}
\bibinfo{author}{\bibfnamefont{G.}~\bibnamefont{Onida}},
  \bibinfo{author}{\bibfnamefont{L.}~\bibnamefont{Reining}}, \bibnamefont{and}
  \bibinfo{author}{\bibfnamefont{A.}~\bibnamefont{Rubio}},
  \bibinfo{journal}{Rev.\ Mod.\ Phys.} \textbf{\bibinfo{volume}{74}},
  \bibinfo{pages}{601} (\bibinfo{year}{2002}).

\bibitem[{\citenamefont{Rieger et~al.}(1999)\citenamefont{Rieger, Steinbeck,
  White, Rojas, and Godby}}]{Rieger}
\bibinfo{author}{\bibfnamefont{M.~M.} \bibnamefont{Rieger}},
  \bibinfo{author}{\bibfnamefont{L.}~\bibnamefont{Steinbeck}},
  \bibinfo{author}{\bibfnamefont{I.~D.} \bibnamefont{White}},
  \bibinfo{author}{\bibfnamefont{H.~N.} \bibnamefont{Rojas}}, \bibnamefont{and}
  \bibinfo{author}{\bibfnamefont{R.~W.} \bibnamefont{Godby}},
  \bibinfo{journal}{Comput.\ Phys.\ Commun.} \textbf{\bibinfo{volume}{117}},
  \bibinfo{pages}{211} (\bibinfo{year}{1999}).

\bibitem[{\citenamefont{Steinbeck et~al.}(2000)\citenamefont{Steinbeck, Rubio,
  Reining, Torrent, White, and Godby}}]{gwstGL}
\bibinfo{author}{\bibfnamefont{L.}~\bibnamefont{Steinbeck}},
  \bibinfo{author}{\bibfnamefont{A.}~\bibnamefont{Rubio}},
  \bibinfo{author}{\bibfnamefont{L.}~\bibnamefont{Reining}},
  \bibinfo{author}{\bibfnamefont{M.}~\bibnamefont{Torrent}},
  \bibinfo{author}{\bibfnamefont{I.~D.} \bibnamefont{White}}, \bibnamefont{and}
  \bibinfo{author}{\bibfnamefont{R.~W.} \bibnamefont{Godby}},
  \bibinfo{journal}{Comput.\ Phys.\ Commun.} \textbf{\bibinfo{volume}{125}},
  \bibinfo{pages}{105} (\bibinfo{year}{2000}).

\bibitem[{\citenamefont{Freysoldt et~al.}(2007)\citenamefont{Freysoldt, Eggert,
  Rinke, Schindlmayr, Godby, and Scheffler}}]{myCPC}
\bibinfo{author}{\bibfnamefont{C.}~\bibnamefont{Freysoldt}},
  \bibinfo{author}{\bibfnamefont{P.}~\bibnamefont{Eggert}},
  \bibinfo{author}{\bibfnamefont{P.}~\bibnamefont{Rinke}},
  \bibinfo{author}{\bibfnamefont{A.}~\bibnamefont{Schindlmayr}},
  \bibinfo{author}{\bibfnamefont{R.~W.} \bibnamefont{Godby}}, \bibnamefont{and}
  \bibinfo{author}{\bibfnamefont{M.}~\bibnamefont{Scheffler}},
  \bibinfo{journal}{Comput. Phys. Commun.} \textbf{\bibinfo{volume}{176}},
  \bibinfo{pages}{1} (\bibinfo{year}{2007}).

\bibitem[{met()}]{metals}
\bibinfo{note}{With the exception of metallic systems at zero frequency, for
  which $P_{\mathbf k}(\mathbf 0, \mathbf 0)$ approaches a finite value as
  $\mathbf k\rightarrow \mathbf 0$. The wings ($\mathbf G= \mathbf 0 \ne
  \mathbf G'$ or $\mathbf G\ne \mathbf 0 = \mathbf G'$) behave
  analogously.\cite{myCPC}}.

\bibitem[{\citenamefont{Pick et~al.}(1970)\citenamefont{Pick, Cohen, and
  Martin}}]{PickCohenMartin}
\bibinfo{author}{\bibfnamefont{R.~M.} \bibnamefont{Pick}},
  \bibinfo{author}{\bibfnamefont{M.~H.} \bibnamefont{Cohen}}, \bibnamefont{and}
  \bibinfo{author}{\bibfnamefont{R.~M.} \bibnamefont{Martin}},
  \bibinfo{journal}{Phys.\ Rev.\ B} \textbf{\bibinfo{volume}{1}},
  \bibinfo{pages}{910} (\bibinfo{year}{1970}).

\bibitem[{\citenamefont{Hott}(1991)}]{Hott}
\bibinfo{author}{\bibfnamefont{R.}~\bibnamefont{Hott}},
  \bibinfo{journal}{Phys.\ Rev.\ B} \textbf{\bibinfo{volume}{44}},
  \bibinfo{pages}{1057} (\bibinfo{year}{1991}).

\bibitem[{\citenamefont{Pulci et~al.}(1998)\citenamefont{Pulci, Onida,
  Del~Sole, and Reining}}]{Pulci}
\bibinfo{author}{\bibfnamefont{O.}~\bibnamefont{Pulci}},
  \bibinfo{author}{\bibfnamefont{G.}~\bibnamefont{Onida}},
  \bibinfo{author}{\bibfnamefont{R.}~\bibnamefont{Del~Sole}}, \bibnamefont{and}
  \bibinfo{author}{\bibfnamefont{L.}~\bibnamefont{Reining}},
  \bibinfo{journal}{Phys.\ Rev.\ Lett.} \textbf{\bibinfo{volume}{81}},
  \bibinfo{pages}{5374} (\bibinfo{year}{1998}).

\bibitem[{\citenamefont{Wenzien et~al.}(1995)\citenamefont{Wenzien, Cappellini,
  and Bechstedt}}]{Wenzien}
\bibinfo{author}{\bibfnamefont{B.}~\bibnamefont{Wenzien}},
  \bibinfo{author}{\bibfnamefont{G.}~\bibnamefont{Cappellini}},
  \bibnamefont{and}
  \bibinfo{author}{\bibfnamefont{F.}~\bibnamefont{Bechstedt}},
  \bibinfo{journal}{Phys.\ Rev.\ B} \textbf{\bibinfo{volume}{51}},
  \bibinfo{pages}{14701} (\bibinfo{year}{1995}).

\bibitem[{rem({\natexlab{a}})}]{remark:gap}
\bibinfo{note}{All slab systems used in this study turn out to have a direct
  gap located at the $\Gamma$ point.}

\bibitem[{\citenamefont{Schindlmayr}(2000)}]{Schindlmayr}
\bibinfo{author}{\bibfnamefont{A.}~\bibnamefont{Schindlmayr}},
  \bibinfo{journal}{Phys. Rev. B} \textbf{\bibinfo{volume}{62}},
  \bibinfo{pages}{12573} (\bibinfo{year}{2000}).

\bibitem[{\citenamefont{Ismail-Beigi and Arias}(1999)}]{IsmailBeigiArias}
\bibinfo{author}{\bibfnamefont{S.}~\bibnamefont{Ismail-Beigi}}
  \bibnamefont{and} \bibinfo{author}{\bibfnamefont{T.~A.} \bibnamefont{Arias}},
  \bibinfo{journal}{Phys. Rev. Lett.} \textbf{\bibinfo{volume}{82}},
  \bibinfo{pages}{2127} (\bibinfo{year}{1999}).

\bibitem[{COH()}]{COHSEX}
\bibinfo{note}{Static COHSEX is derived from the COHSEX partitioning of the
  non-self-consistent $G_0W_0$ self-energy assuming that the energy-dependence
  of the screened interaction can be neglected \cite{Aulbur}}.

\bibitem[{\citenamefont{Delerue et~al.}(2003)\citenamefont{Delerue, Allan, and
  Lannoo}}]{Delerue}
\bibinfo{author}{\bibfnamefont{C.}~\bibnamefont{Delerue}},
  \bibinfo{author}{\bibfnamefont{G.}~\bibnamefont{Allan}}, \bibnamefont{and}
  \bibinfo{author}{\bibfnamefont{M.}~\bibnamefont{Lannoo}},
  \bibinfo{journal}{Phys. Rev. Lett.} \textbf{\bibinfo{volume}{90}},
  \bibinfo{pages}{076803} (\bibinfo{year}{2003}).

\bibitem[{rem({\natexlab{b}})}]{remark:s2vrule}
\bibinfo{note}{As a rule of thumb, we found that in the relevant range of slab
  and vacuum thicknesses ($s$ and $v$, respectively), the finite-vacuum effect
  depends mainly on $s+2v$.}

\bibitem[{\citenamefont{Jackson}(1975)}]{Jackson}
\bibinfo{author}{\bibfnamefont{J.~D.} \bibnamefont{Jackson}},
  \emph{\bibinfo{title}{Classical Electrodynamics}} (\bibinfo{publisher}{Wiley,
  New York}, \bibinfo{year}{1975}).

\bibitem[{eps()}]{eps_profile}
\bibinfo{note}{Even a continuous dielectric profile $\varepsilon(z)$ can be
  approximated by such a layer model when the profile is discretized into
  individual layers of a sufficiently small thickness. However, the profiles
  used for this work are step-like functions.}

\end{thebibliography}

\end{document}